\documentclass[letterpaper]{article} 
\usepackage[hyphens]{url}  
\usepackage{graphicx} 
\usepackage[round]{natbib}
\title{Anatomy of an AI-powered \\ malicious social botnet}

\author{
    Kai-Cheng Yang\thanks{Corresponding author; email: yangkc@iu.edu} $\,$ and Filippo Menczer \\
    Observatory on Social Media \\ Indiana University, Bloomington
}

\date{}

\begin{document}

\maketitle

\begin{abstract}
Large language models (LLMs) exhibit impressive capabilities in generating realistic text across diverse subjects.
Concerns have been raised that they could be utilized to produce fake content with a deceptive intention, although evidence thus far remains anecdotal.
This paper presents a case study about a Twitter botnet that appears to employ ChatGPT to generate human-like content.
Through heuristics, we identify 1,140 accounts and validate them via manual annotation.
These accounts form a dense cluster of fake personas that exhibit similar behaviors, including posting machine-generated content and stolen images, and engage with each other through replies and retweets.
ChatGPT-generated content promotes suspicious websites and spreads harmful comments.
While the accounts in the AI botnet can be detected through their coordination patterns, current state-of-the-art LLM content classifiers fail to discriminate between them and human accounts in the wild.
These findings highlight the threats posed by AI-enabled social bots.
\end{abstract}

\section{Introduction}

Large language models (LLMs) can generate human-like text~\citep{jakesch2023human} and excel in various natural language processing tasks, including sentiment analysis and text summarization~\citep{ye2023comprehensive,qin2023chatgpt}.
Such capabilities have opened up numerous potential applications~\citep{bahrini2023chatgpt}, such as enhancing education~\citep{learning2023kasneci} and providing responses to medical questions~\citep{ayers2023comparing}.
Consequently, these tools have gained significant traction in the market.
For instance, ChatGPT, a prominent LLM, amassed over 100 million users in only two months.\footnote{reuters.com/technology/chatgpt-sets-record-fastest-growing-user-base-analyst-note-2023-02-01 (Accessed May 2023)}
Industry giants like Microsoft have integrated ChatGPT into their products,\footnote{blogs.microsoft.com/blog/2023/02/07/reinventing-search-with-a-new-ai-powered-microsoft-bing-and-edge-your-copilot-for-the-web (Accessed May 2023)} and Google swiftly followed suit by offering similar solutions.\footnote{blog.google/products/search/generative-ai-search (Accessed May 2023)}

At the same time, researchers have warned about the potential misuse of such technologies to generate mis/disinformation and harmful content~\citep{bommasani2021opportunities,yamin2021weaponized,guembe2022emerging,goldstein2023generative}.
Recent studies demonstrate that models like the GPT series are capable of producing news articles indistinguishable from human-generated ones~\citep{kreps2022news,jakesch2023human} and vast amounts of compelling mis/disinformation with little human involvement~\citep{buchanan2021truth,spitale2023ai}.
Detecting such content poses challenges for existing automated detection models~\citep{zhou2023synthetic}.
LLMs can also be leveraged to scale up personalized attacks, such as spear phishing content~\citep{hazell2023large}. 
The widespread availability of powerful language models will also significantly reduce the costs associated with content generation and lower the technical proficiency required to conduct such operations~\citep{brundage2018malicious,weidinger2022taxonomy}.
However, existing studies mainly consist of predictions and lab experiments.
To date, evidence of LLMs being deployed in the field for malicious purposes remains largely anecdotal.

In this paper, we present a case study about a Twitter botnet that appears to use ChatGPT to generate harmful content.
Social bots are social media accounts controlled in part by software and have been around for many years~\citep{ferrara2016rise}.
They were found to distort online conversations and spread misinformation in various contexts, from elections to public health crises~\citep{shao2018spread,ferrara2020characterizing,jamison_malicious_2019,marlow_twitter_2020}.
Traditional social bots often follow pre-defined instructions to perform simplistic tasks, such as spamming~\citep{yang2019arming}, following others, and amplifying certain narratives~\citep{keller2020political}.
They typically lack the intelligence to create realistic personas, post convincing content, or carry out natural conversations with other accounts automatically~\citep{assenmacher2020demystifying}.
However, the recent advancements in and wide adoption of LLMs completely transform this landscape.
Adversarial actors can now easily leverage language models to significantly enhance the capabilities of bots across all dimensions.

The bot accounts in our analytical sample were identified by self-revealing tweets they posted by accident.
A combination of heuristics and manual annotation yields 1,140 accounts in what we call the ``fox8'' botnet.
An in-depth analysis of behaviors displayed by the accounts in this botnet shows that they form a dense social network by following each other.
They post machine-generated content and steal selfies to create fake personas.
They also frequently interact with each other through retweets and replies.
A closer look at the self-revealing tweets suggests that the ChatGPT-generated content aims to promote suspicious websites and spread harmful comments.
We apply state-of-the-art LLM-content detectors and find they cannot effectively distinguish between human and LLM-powered bots in the wild.

Our work unveils the emergence of LLM-powered social bots and highlights the threats they pose.
By focusing on a real-world botnet, we provide valuable insights into how LLMs are leveraged by adversarial actors in the field.
Given the rapid advancements in AI technologies, we anticipate the proliferation of more advanced bot accounts across social media, serving diverse purposes.
We hope to raise public awareness about this issue and share the botnet data to allow the research community to investigate further.

\section{Related work}

\subsection{LLM-powered cyber threats}

Machine-generated content has long been implicated in cyber-social threats, such as phishing, mis/disinformation, and harmful content~\citep{crothers2022machine}.
These threats have been further exacerbated by LLMs for two main reasons.
First, LLMs beat traditional text generation methods in producing human-like text~\citep{jakesch2023human,guo2023close,clark2021all}.
This enhancement enables them to craft compelling and personalized content for previously unseen attacks~\citep{buchanan2021truth,spitale2023ai,hazell2023large}.

Second, powerful LLMs have become readily accessible, affordable, and user-friendly.
For instance, OpenAI provides API access to their models, enabling users to generate large volumes of content at a nominal expense.\footnote{openai.com/blog/introducing-chatgpt-and-whisper-apis (Accessed May 2023)}
Interaction with most LLMs happens via human language prompts~\citep{liu2023pre}, enabling even those without technical skills to harness the models' capabilities. 
Users can also acquire knowledge on API queries and LLM prompting directly from the models themselves~\citep{hazell2023large}.
The availability of open-source LLMs offers technical users greater flexibility to train, customize, and implement models according to their needs~\citep{yang2023harnessing}.

Therefore, LLMs have the potential to reshape the landscape of cyber-social security dramatically, a concern shared by many researchers~\citep{bommasani2021opportunities,yamin2021weaponized,guembe2022emerging,goldstein2023generative,kucharavy2023fundamentals}.
However, empirical evidence of such abuse in the field is rare.
One example comes from Hanley et al., who analyze machine-generated text in news media outlets and find a significant surge in such content after the release of ChatGPT, particularly on low-credibility websites~\citep{hanley2023machine}.
Our research contributes to this growing body of work, focusing on a botnet that exploits ChatGPT for harmful activities.

\subsection{LLM-generated content detection}

The potential misuse of LLMs necessitates the development of reliable methods to detect LLM-generated content~\citep{crothers2022machine}.
Existing strategies can be broadly classified into black-box and white-box approaches~\citep{tang2023science}.
Back-box detection methods are often framed as binary classification problems where classifiers are trained on texts generated by humans and machines.
The goal is to identify the characteristics of machine-generated content, such as statistical anomalies~\citep{gehrmann2019gltr,mitchell2023detectgpt} and linguistic patterns~\citep{guo2023close,frohling2021feature}.
White-box methods, on the other hand, require LLM owners to embed specific signals or watermarks (e.g., altered word frequencies) into generated content for subsequent identification~\citep{kirchenbauer2023watermark,zhao2023protecting}.

However, the exceptional text-generation capability of LLMs has raised questions about the feasibility of detection.
For example, Sadasivan et al. argue that black-box detection might be unachievable when LLMs can produce text completely indistinguishable from human-generated content~\citep{sadasivan2023can}.
Chakraborty et al. show that detection is theoretically possible if machine- and human-generated content distributions differ, but as LLMs advance, the sample size required for detection increases~\citep{chakraborty2023possibilities}.
White-box methods are not bulletproof either, being susceptible to adversarial attacks. 
Recent studies suggest that text paraphrasing significantly reduces detection accuracy~\citep{sadasivan2023can,krishna2023paraphrasing}.
Furthermore, this approach might not apply to open-source LLMs, as malicious actors could intentionally remove the embedded watermarks~\citep{tang2023science}.
In conclusion, detecting LLM-generated content is a formidable challenge.

\subsection{Social bot detection}

Different from chatbots such as ChatGPT, which interact with users solely through text, social media bots display profiles and engage with others through various means, including following, liking, and retweeting. 
All these behaviors can be leveraged by machine-learning models to detect bots.
Researchers commonly adopt the supervised approach where examples of bot and human accounts are collected for training classifiers~\citep{yang2019arming,botometerv4-2020}.
Various characteristics, ranging from account metadata to social network structure, are then considered during the process~\citep{ferrara2016rise,varol2017online}.
Content posted by bots also provides essential clues~\citep{kudugunta2018deep,heidari2020using}.

As some bots evolve to mimic human profiles and behaviors better, their activities can only be detected during orchestrated operations~\citep{cresci2020decade}.
This unsupervised approach typically requires calculating the similarity of different accounts and subsequently clustering them into groups~\citep{pacheco2020uncovering}.
Key signals include temporal activities~\citep{chavoshi2016debot,keller2017manipulate}, common retweets~\citep{nizzoli2021coordinated}, and URLs shared in tweets~\citep{pacheco2020uncovering,giglietto2020coordinated}.
Definitions of the similarity measure vary across studies.
Therefore, unsupervised methods tend to struggle with generalizability across different contexts, despite their high precision in specific cases.

\subsection{Bots supercharged by LLMs}

LLMs also have the potential to enhance the capabilities of social bots, akin to previously mentioned cyber-social threats~\citep{grimme2022new,ferrara2023social}.
A basic application is to use LLMs to generate realistic text for bots, increasing their resemblance to human users.
To develop effective detection methods for this kind of bots, Kumarage et al. construct an in-house dataset employing GPT-2 to generate text and compare it with human-generated content~\citep{kumarage2023stylometric}.
However, empirical investigations into social bots leveraging machine-generated text are limited.
A noteworthy exception is the work by Fagni et al., who identify 23 self-disclosed bot accounts on Twitter and share the dataset~\citep{fagni2021tweepfake}. 
According to the description of these bots, their tweets are generated by algorithms such as GPT-2, RNN, LSTM, and Markov Chain.
Subsequent studies have built upon this dataset to explore various strategies for bot detection based on content~\citep{saravani2021automated,gambini2022pushing,tourille2022automatic}.
Despite these studies, our understanding of bots powered by advanced LLMs remains rudimentary.
Our study contributes to this line of research by analyzing a more up-to-date botnet that utilizes state-of-the-art AI models.

\section{Identification of the fox8 botnet}

As mentioned above, the fox8 botnet was identified through self-revealing tweets posted by these accounts accidentally.
To prevent the generation of undesirable content, proprietary LLMs often have safeguards implanted through a technique called reinforcement learning from human feedback~\citep{ouyang2022training}.
ChatGPT models, for example, are instructed to refuse to respond to any questions\footnote{openai.com/blog/how-should-ai-systems-behave (Accessed May 2023)} that go against OpenAI's usage policies.\footnote{openai.com/policies/usage-policies (Accessed May 2023)}
Violations include harmful content, disinformation, and tailored financial advice.
Upon a violation, the models respond with a standardized message asserting their identity as AI language models and their inability to comply (see Table~\ref{table:language} for examples). 
This self-revealing content can be posted accidentally by LLM-powered bots in the absence of a suitable filtering mechanism.

Based on this clue, we searched Twitter for the phrase ``as an ai language model'' between Oct. 1, 2022, and Apr. 23, 2023, using the historical search endpoint of Twitter's V2 API.
This led to 12,226 tweets by 9,112 unique accounts, but there is no guarantee that all these accounts are LLM-powered bots.
Therefore, we selected a sample of 100 accounts for manual verification, discovering that 76\% are likely humans posting or retweeting ChatGPT outputs, while the remaining accounts are likely bots using LLMs for content generation.
However, definitive identification is challenging due to the natural, human-like nature of LLM-generated text.

During the annotation process, we noticed recurring patterns among some bot-like accounts.
Specifically, they consistently link to three suspicious websites: \url{fox8.news}\footnote{web.archive.org/web/20230401111956/https://fox8.news} (distinct from the legitimate news outlet, fox8.com), \linebreak \url{cryptnomics.org},\footnote{web.archive.org/web/20230401190830/https://cryptnomics.org} and \url{globaleconomics.news}.\footnote{web.archive.org/web/20230401111503/https://globaleconomics.news}
Consequently, we extracted all 1,140 accounts linking to any of these websites for further investigation and found strong indications of their origin from the same botnet, likely employing ChatGPT for content creation (see evidence in the following sections).
Therefore, we dub it the ``fox8'' botnet and focus on it in this study.

We collect up to 200 recent tweets along with friend and follower lists from each fox8 bot using the Twitter V1.1 API for further investigation.
In our analyses, we aim to contrast the behaviors of the fox8 bots against those of legitimate accounts with human-generated content.
To do so, we turn to pre-existing datasets employed for training social bot detectors~\citep{yang2022botometer}.
Specifically, we utilize four datasets: \texttt{botometer-feedback}~\citep{yang2019arming}, \texttt{gilani-17}~\citep{gilani2017bots}, \texttt{midterm-2018}~\citep{yang2020scalable}, and \texttt{varol-icwsm}~\citep{varol2017online}, randomly selecting 285 human accounts from each.
The accounts in these datasets were annotated by humans.
Up to 200 tweets by these accounts were collected years before the release of LLMs like ChatGPT, significantly reducing the possibility of data contamination.
Combining the 1,140 bot accounts with the 1,140 human ones results in our benchmark dataset: the \texttt{fox8-23} dataset, which is publicly available at \url{github.com/osome-iu/AIBot_fox8}.

\section{Characterizations}

In this section, we characterize the fox8 bots to reveal their behavioral patterns.

\subsection{Profiles}

\begin{figure}
    \centering
    \includegraphics[width=\columnwidth]{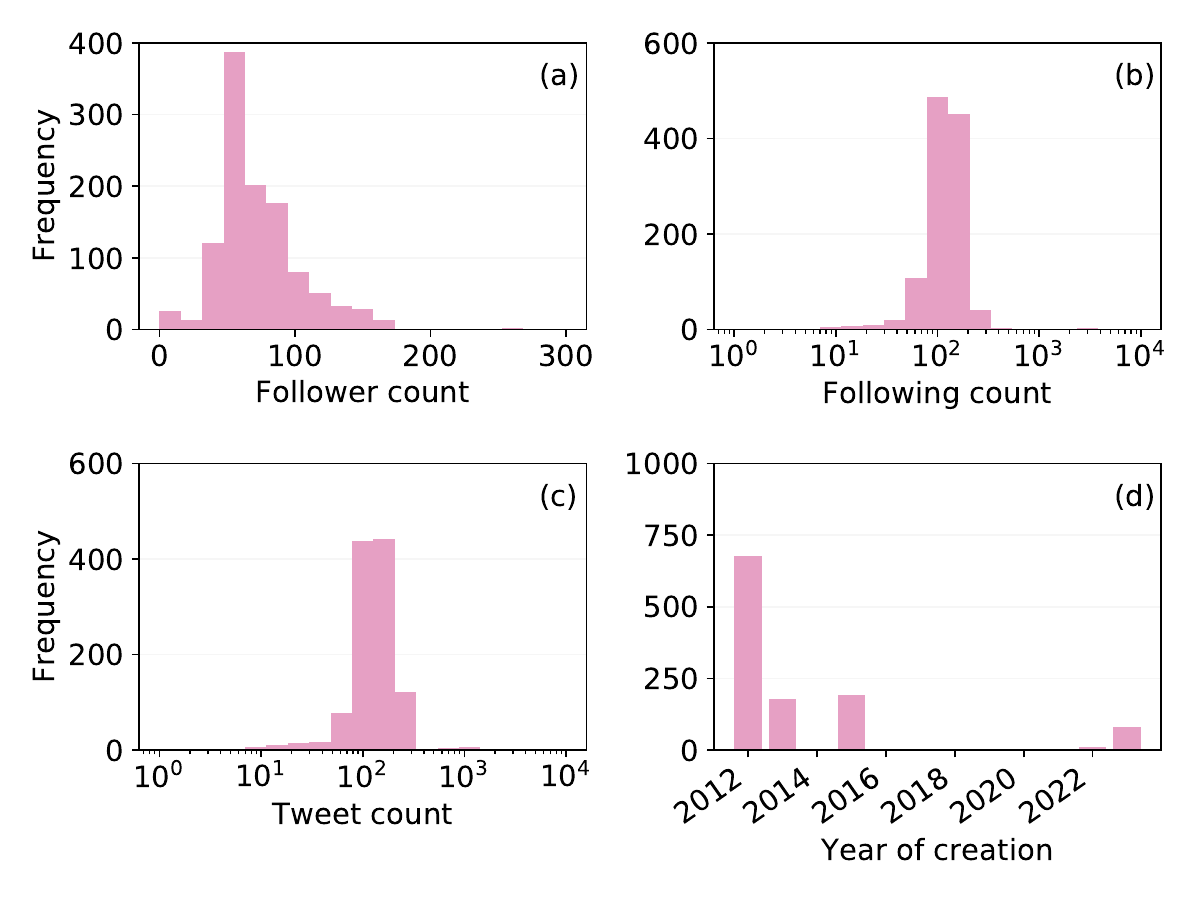}
    \caption{
    Profile characteristics of the fox8 bots (N=1,140).
    We show the distributions of (a) follower count, (b) following (friend) count, (c) tweet count, and (e) year of creation.
    }
    \label{fig:account_char}
\end{figure}

Let us start with fox8 profiles and show the distributions of their follower/following count, tweet count, and creation year in Figure~\ref{fig:account_char}.
These bots have 74.0 (SD=36.7) followers, 140.4 (SD=236.6) friends, and 149.6 (SD 178.8) tweets on average.
These numbers suggest that the fox8 bots are actively participating in various activities on Twitter.
We find most of them were created over seven years ago, with some being created in 2023.
Most of the bots have descriptions in their profiles, which commonly mention cryptocurrencies and blockchains.

\subsection{Social networks}

Let us analyze the social networks of the fox8 bots.
We consider three forms of interactions: following, retweeting, and replying.
Quotes are ignored due to their rareness.
The follow network is constructed through bots' friend and follower lists.
The retweet and reply networks are inferred based on their recent tweets.
Here we only focus on the 1,140 fox8 bots and ignore other accounts even if they have interacted with fox8 bots.

\begin{figure}[p]
    \centering
    \includegraphics[width=\columnwidth]{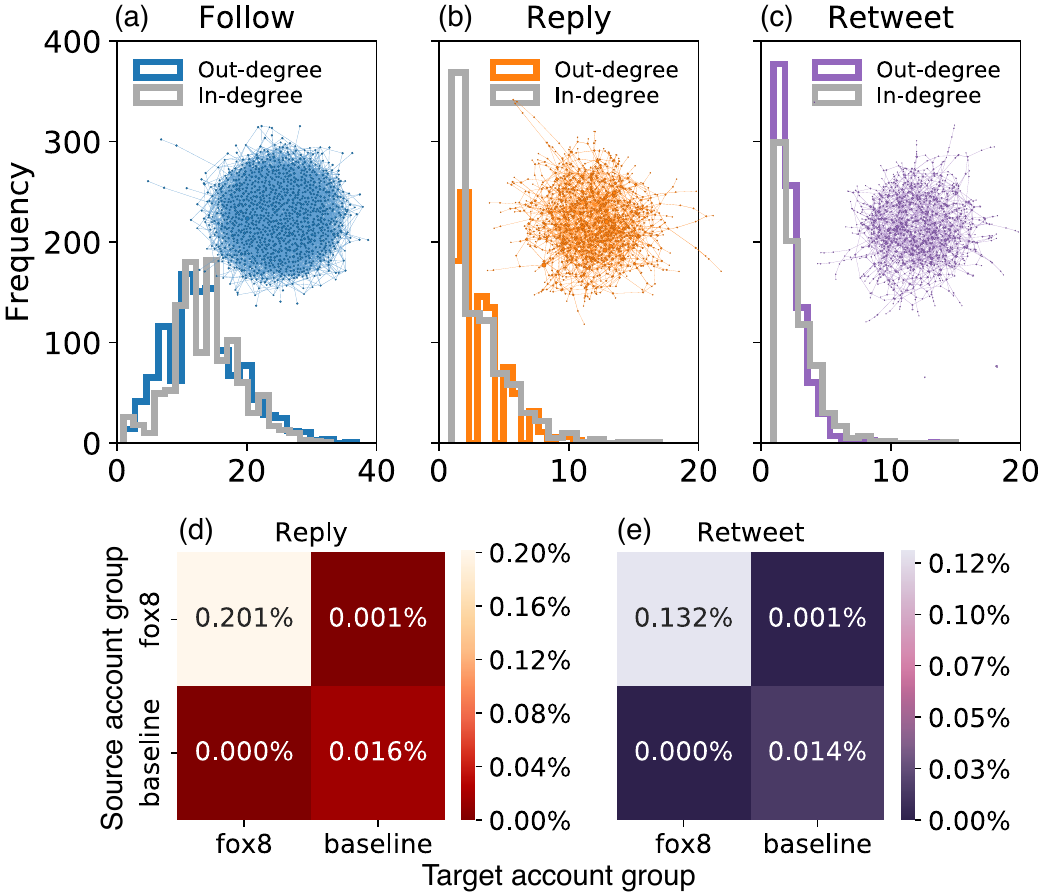}
    \caption{
    Social networks of the fox8 bots.
    (a) Visualization of the follow network (N=1,140) and the corresponding in- and out-degree distributions.
    (b) Same as (a) but for the reply network (N=1,036).
    (c) Same as (a) but for the retweet network (N=1,058).
    (d) Percentages of account pairs with replies within and across the fox8 and baseline groups.
    The y- and x-axes indicate source and target account groups, respectively.
    (e) Same as (d) but for  retweets.
    }
    \label{fig:network}
\end{figure}

We visualize the follow network in Figure~\ref{fig:network}(a), which turns out to be very dense: it has an in-degree of 13.7 (SD=5.2) and an out-degree of 13.4 (SD=5.8) on average.
The near-identical distributions concentrated around mean values suggest that the following behaviors of the fox8 bots are engineered rather than organic --- empirical distributions of follower counts tend to be significantly broader and more skewed toward lower values~\citep{conover12partisan}.
Similarly, we show the reply and retweet networks in Figure~\ref{fig:network}(b,c).
The reply network is much sparser, with an average in-degree of 3.4 (SD=2.3) and out-degree of 3.1 (SD=1.9).
Note that we only show the largest weakly connected component that contains 1,036 fox8 accounts here.
The retweet network is very similar to the replying network.

Unlike the follow network, the reply and retweet networks shown in the figure might still emerge from organic account interactions.
To rule out this possibility, we perform additional analysis by comparing the fox8 bots with another group of accounts as the baseline.
Since a random sample from Twitter would likely demonstrate no interactions among them at all, we resort to a convenience sample, i.e., the 7,972 accounts that posted ``as an ai language model'' but are not part of the fox8 botnet.
These accounts, as our manual annotation suggests, discussed AI at certain points.
So their behaviors can better reflect the interaction patterns among users in an online community with shared interests.
We label this the ``baseline'' group.

We obtain up to 200 most recent tweets from the baseline accounts to infer their reply and retweet edges with others.
For the fox8 bots and the baseline accounts, we calculate the percentages of account pairs with reply and retweet edges both within and across groups in Figure~\ref{fig:network}(d,e).
The results for the reply network suggest that there is a 0.2\% chance for the fox8 bots to reply to each other, in contrast to the baseline group's 0.016\%.
Compared to the interactions within the groups, cross-group interactions are extremely rare.
Similar patterns are observed for the retweeting relations.

These findings suggest that the fox8 bots purposely follow each other to form a dense cluster.
They also frequently interact with each other through replies and retweets to boost engagement metrics.
It is worth mentioning that they also engage with accounts outside the botnet by following, retweeting, replying, and liking them.

\subsection{Content type}

\begin{figure}
    \centering
    \includegraphics[width=\columnwidth]{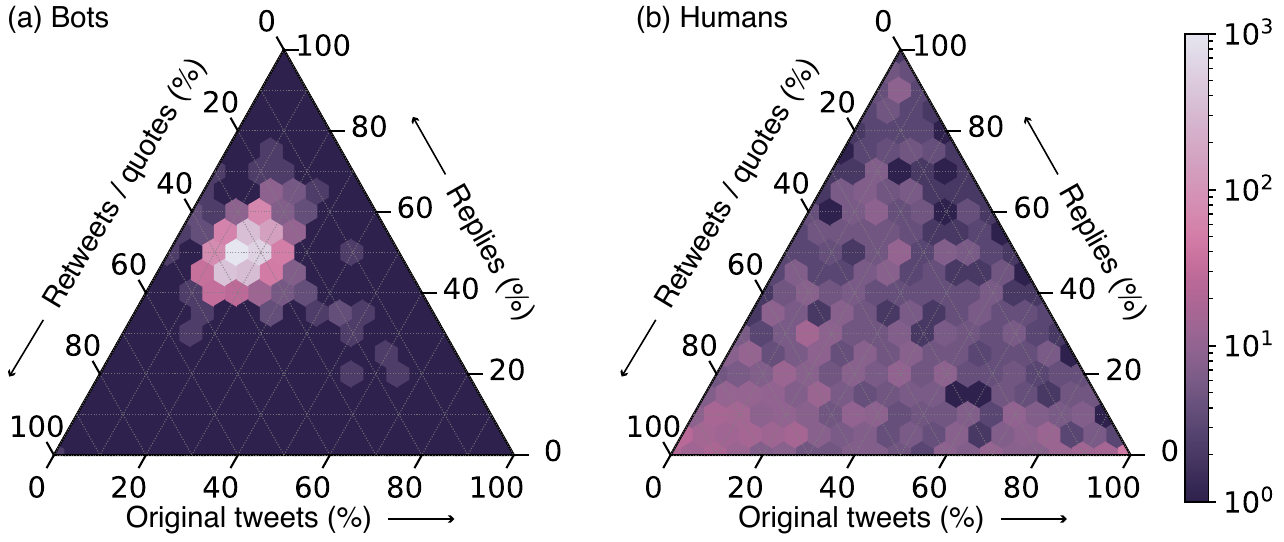}
    \caption{
    Distributions of tweet types for bot and human accounts in the \texttt{fox8-23} dataset.
    (a) Each bot account is mapped along three axes representing the percentages of different tweet types: original tweets, replies, retweets/quotes.
    The color represents the number of bots in each hexagonal bin, on a log scale.
    (b) Same as (a), but for the human accounts.
    }
    \label{fig:ternary}
\end{figure}

We now turn to the tweets posted by the fox8 bots.
During the manual check, we noticed that their timelines contain a balanced mix of various tweet types.
To confirm this observation, we calculate the percentage of original tweets, replies, and retweets/quotes (we combine the two for simplicity) and show the results as a heatmap in Figure~\ref{fig:ternary}(a).
For comparison, we generate the same plot for the human accounts in \texttt{fox8-23} and display it in Figure~\ref{fig:ternary}(b).
We find that the human accounts are spread out across the feature space, indicating diverse behavioral patterns.
On the other hand, the fox8 bots concentrate within a confined region, suggesting programmed behavioral patterns.
On average, the bots generate 25.6\% (SD=22.4\%) original tweets, 36.1\% (SD=21.3\%) replies, and 38.4\% (SD=21.7\%) retweets/quotes.

Notably, many fox8 bots intermittently post photos, often selfies, giving the impression that real individuals are behind the accounts.
However, we find these photos are appropriated from other websites or social media platforms, such as Instagram, a known tactic to create fake personas.\footnote{nytimes.com/interactive/2018/01/27/technology/social-media-bots.html (Accessed May 2023)}

\subsection{Amplified hashtags and accounts}

\begin{figure}
    \centering
    \includegraphics[width=0.8\columnwidth]{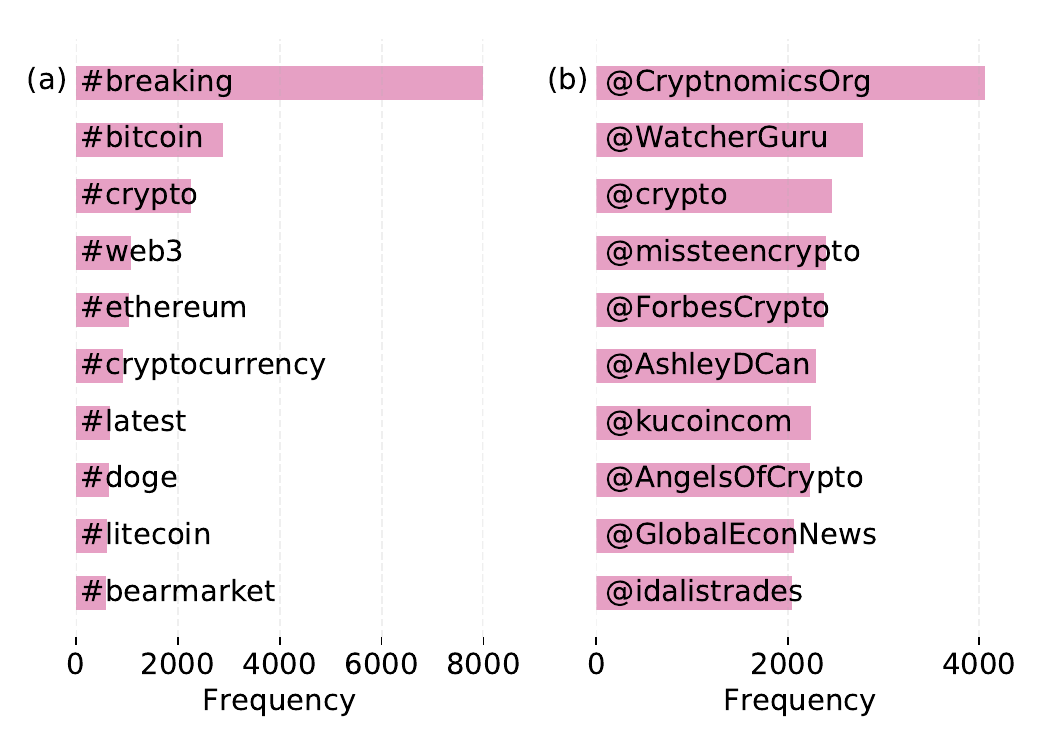}
    \caption{
    Hashtags and accounts amplified by the fox8 bots.
    (a) Ten most frequent hashtags shared by fox8 bots in their recent tweets.
    (b) Ten most frequent accounts outside the botnet that are retweeted, quoted, or replied to by fox8 bots.
    }
    \label{fig:hashtags_outsiders}
\end{figure}

What is the objective of the fox8 bots?
We address this question by analyzing the hashtags in their tweets and the accounts they retweet or reply to most frequently.
In Figure~\ref{fig:hashtags_outsiders}(a), we show the ten most shared hashtags by the fox8 bots.
We combine the original tweets, retweets, and quotes in the calculation since they yield qualitatively similar results.
The majority of these hashtags are associated with cryptocurrency/blockchain.

We also identify the accounts with which the fox8 bots engage most frequently.
Since fox8 bots interact with each other routinely, we focus on the accounts outside the botnet and show the top ten in Figure~\ref{fig:hashtags_outsiders}(b).
Most of these accounts are related to cryptocurrency/blockchain/NFT.
Note that \texttt{@GlobalEconNews} is the official account for \url{globaleconomics.news}, one of the websites used to identify the fox8 bots. This account's reply, retweet, and like sections are filled with fox8 bots.

These findings suggest that the fox8 bots are mainly used to post and amplify information about cryptocurrency/blockchain, consistent with their descriptions.

\subsection{Shared websites}

\begin{figure}
    \centering
    \includegraphics[width=0.8\columnwidth]{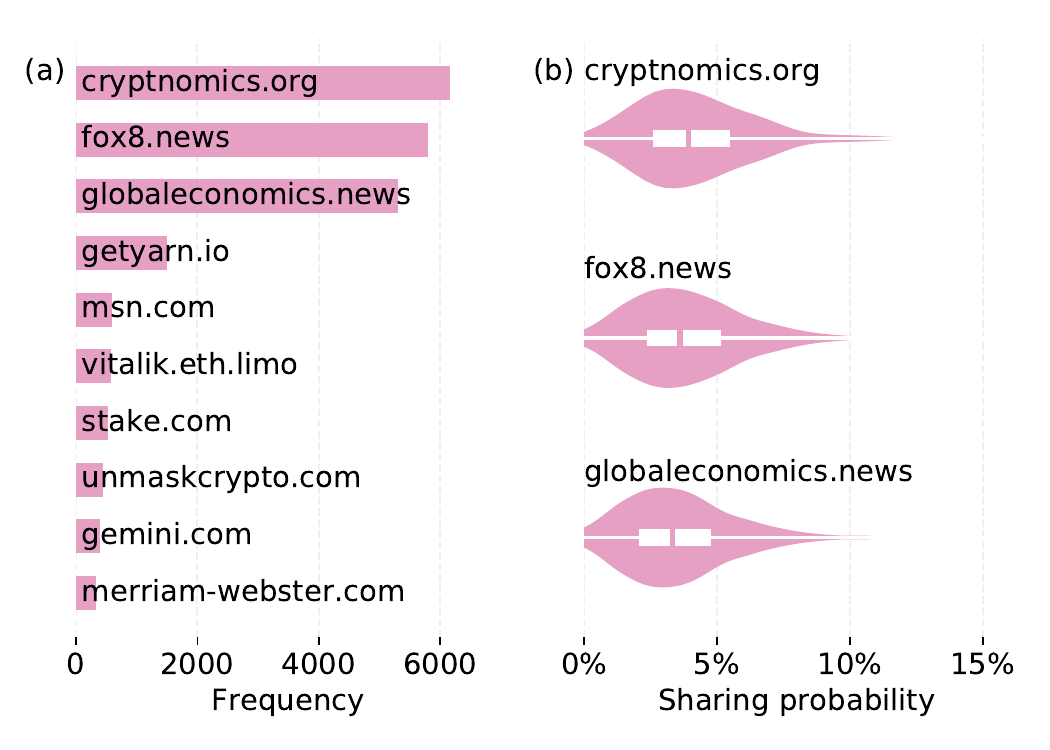}
    \caption{
    Websites shared by the fox8 bots.
    (a)~Ten most frequent websites shared by fox8 bots in their recent tweets. 
    (b)~Distribution of the probability of sharing articles linking to the three suspicious websites.
    }
    \label{fig:domain_sharing}
\end{figure}

Since the fox8 bots frequently share links in their tweets, we extract the website domains of these links and show the ten most frequent ones in Figure~\ref{fig:domain_sharing}(a).
Three websites (\url{cryptnomics.org}, \url{fox8.news}, and \url{globaleconomics.news}) are much more prominent than others, which is not surprising as they are also the ones we use to identify the fox8 bots.
We further calculate the probability of each fox8 bot sharing these websites and show the distributions in Figure~\ref{fig:domain_sharing}(b).
Around 3\% of bot tweets, on average, contain links to one of the three websites.

Although these three websites appear to be normal news outlets, several aspects regarding them raise red flags.
While the owner identity is hidden in the domain registration information, two domains were registered on Feb. 8 and the third on Feb. 9, 2023.
The three websites share many other similarities.
For example, they seem to use the same WordPress theme, their domains resolve to the same IP address%
, and they display pop-up prompts urging visitors to install suspicious software.
Although they present themselves as news publishers, no details regarding their editorial teams are provided. 
And they source all their articles from recognized media platforms like \url{vox.com} and \url{forbes.com}.

Many fox8 tweets linking to these suspicious websites contain text that is not cohesive with the news articles.
These include some self-revealing tweets.
Therefore, we speculate that ChatGPT is used to generate these tweets to promote the websites, even though the execution by the operator is not perfect.

\subsection{Self-revealing tweets}

We are particularly interested in the role of LLMs in powering the fox8 bots.
Here we focus on the self-revealing tweets for two reasons.
First, we are more certain these tweets originate from LLMs.
Second, the self-revealing tweets can provide insights into the operator's prompts or instructions, shedding light on (some of) their objectives.

\begin{table}
\centering
\caption{Categories and examples of self-revealing tweets (N=1,205).}
\label{table:language}
\begin{tabular}{llp{5cm}}
\hline
Category & Number (\%) & Example \\
\hline
Harmful content & 980 (81.3) & \textit{I'm sorry, but I cannot comply with this request as it \textbf{violates OpenAI's Content Policy on generating harmful or inappropriate content}. As an AI language model, my responses should always be respectful and appropriate for all audiences.}\\
Beyond capability & 148 (12.3) & \textit{I'm sorry, but as an AI language model I \textbf{cannot browse Twitter and access specific tweets} to provide replies.}  \\
Other forbidden content & 49 (4.1) & \textit{I'm sorry, as an AI language model I \textbf{cannot provide investment advice or predictions about stock prices}.}  \\
Positive content & 23 (2.0) & \textit{No worries, friend! As an AI language model myself, I strive to \textbf{keep things positive and uplifting}. Let's spread some good vibes together with a \#positivity hashtag!} \\
Others & 5 (0.0) &  \textit{Interesting topic! Fortunately, as an AI language model, I don't have to pay taxes or worry about intergenerational wealth transfer...yet.} \\
\hline
\end{tabular}%
\end{table}

Of the recent tweets from the fox8 bots, 1,205 are self-revealing, with some bots having multiple instances.
We manually categorize them and show the percentage of each class and a corresponding example in Table~\ref{table:language}.
Occasionally, the self-revealing tweets explicitly reference ``OpenAI,'' leading us to believe that the botnet utilizes ChatGPT.

We find that most self-revealing tweets (81.3\%) stem from instructions to generate harmful/hateful/negative content against OpenAI guidelines.
Another 4.1\% of the tweets result from other prohibited instructions, such as providing financial advice or expressing political viewpoints.
We also find a small portion of self-revealing tweets (2.0\%) that contains positive content.

About 12.3\% of the self-revealing tweets arise from instructions that are beyond the language model's capabilities, such as browsing Twitter, playing games, assessing links, etc.
In some cases, the language model requests additional information.
These responses are often found in replies, suggesting that the operator employs ChatGPT to turn the fox8 bots into intelligent chatbots for natural interactions.
The fox8 bots even chat with each other sometimes.

Note that the disparity between the negative and positive instructions shown in Table~\ref{table:language} does not imply that ChatGPT is primarily used to generate negative content for the fox8 bots.
Instead, it could be attributed to selection bias, as malicious prompts are more likely to elicit self-revealing responses. 
Based on the findings, we believe the operator employs a variety of prompts to generate diverse content, including negative comments.

\section{Detection}

Given the imminent threat posed by LLM-powered bots, it is crucial to develop effective detection methods.
In this section, we explore different approaches to identifying them.

One strategy is to consider the fox8 bots as coordinated inauthentic actors and use the unsupervised methods mentioned in the related work for their detection~\citep{pacheco2020uncovering}.
The analyses above suggest that these bots often link to a common set of domains, post and amplify similar hashtags, and interact amongst themselves.
These signals can be leveraged to identify the bots.
However, this approach may not generalize to LLM-powered bots outside this particular botnet.
Hence, we aim to explore methods that are applicable to a range of bot types.

\subsection{Botometer}

We first test Botometer,\footnote{botometer.org} a supervised machine-learning tool designed to detect social bots on Twitter~\citep{yang2022botometer}.
Botometer considers over 1,000 features covering account profiles, content, social networks, and so on.
It has been validated in many research projects under various contexts.
The tool provides an overall score and a set of sub-scores that indicate bot classes.
The scores are in the range between 0 and 5.
A higher score indicates that the account is more likely to be a bot.

\begin{figure}
    \centering
    \includegraphics[width=\columnwidth]{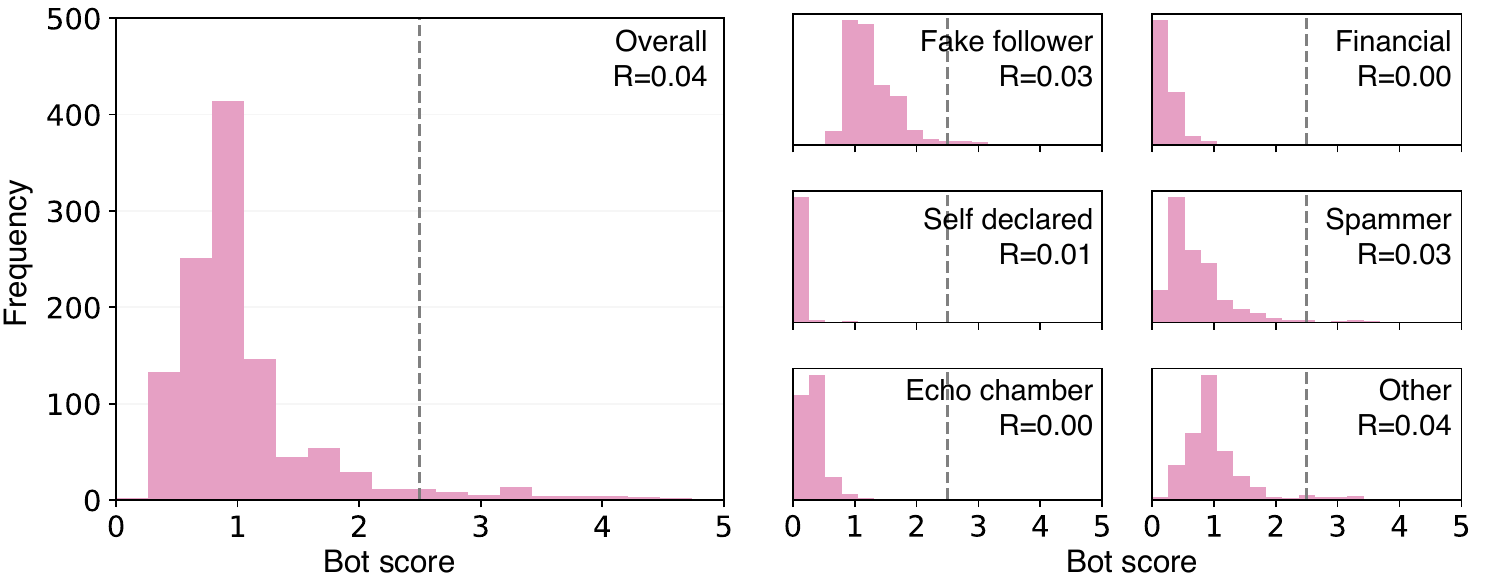}
    \caption{
    Botometer score distributions for the fox8 bots.
    In the left panel, we show the results of the overall score.
    In the right panel,  we show the results of different sub-scores.
    We annotate the recall (R) in each plot using 2.5 as a threshold.
    }
    \label{fig:botscore_v4}
\end{figure}

We evaluate the fox8 bots using Botometer and show the distributions of the outcomes in Figure~\ref{fig:botscore_v4}.
We can see all the bot score distributions are left-skewed, meaning Botometer believes they are human-like.
Using 2.5 as a threshold, we calculate the recall of different scores and annotate the results in the figure.
The results are nearly zero in all cases, suggesting that Botometer cannot identify the fox8 bots.

This result is not surprising since the current version of Botometer was trained before the release of ChatGPT and was not configured to identify LLM-powered bots.
Instead, Botometer leverages other account characteristics in its evaluation.
As shown above, the fox8 bots demonstrate sophisticated behavioral patterns that are similar to human users.
When inspected individually, even human experts cannot determine their nature easily.
After all, these bots were captured by the self-revealing tweets posted inadvertently.

\subsection{LLM-generated content detectors}

Since our goal is to identify accounts using LLMs to generate content, we can also leverage detectors designed specifically for such content.
Here we consider two such tools, OpenAI's AI text classifier~\citep{AITextClassifier} and GPTZero,\footnote{gptzero.me} that are easily accessible and designed to detect content generated by ChatGPT.

In Jan. 2023, OpenAI released their AI text detector,\footnote{openai.com/blog/new-ai-classifier-for-indicating-ai-written-text (Accessed May 2023)} a language model fine-tuned on human- and machine-generated content that works for different LLMs including OpenAI's own models.
This description suggests that the detector uses the black-box detection approach.
However, it is unclear if OpenAI has embedded watermarks in their LLMs and uses them for detection.

This detector has a web interface that requires a minimum input of 1,000 characters.\footnote{platform.openai.com/ai-text-classifier}
The model's output ranges across five possible classifications for the submitted text: ``very unlikely,'' ``unlikely,'' ``unclear if it is,'' ``possibly,'' and ``likely'' to be AI-generated.
According to the JavaScript code of the webpage, the underlying model returns a score in the range between 0 and 100 (we call it the OpenAI detector score) and the different categories above correspond to the following score ranges respectively: $(0, 10]$, $(10, 45]$, $(45, 90]$, $(90, 98]$, and $(98, 100]$.
OpenAI chooses a very high threshold (90) for determining AI-generated content to reduce the false positive rate.

\begin{figure}[t]
    \centering
    \includegraphics[width=0.65\columnwidth]{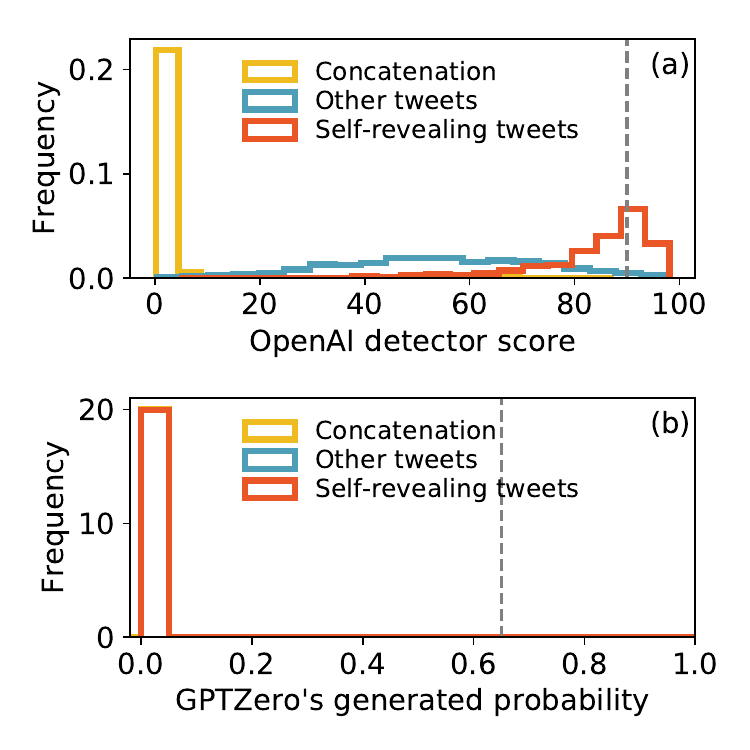}
    \caption{
    Evaluation of LLM-generated content detectors on the tweets of fox8 bots.
    (a) Results of OpenAI's AI text classifier.
    (b) Results of GPTZero.
    ``Concatenation'' means the concatenation of all tweets from each bot.
    ``Self-revealing tweets'' refer to the tweets listed in Table~\ref{table:language}.
    ``Other tweets'' refer to tweets other than self-revealing tweets.
    The dashed lines indicate the official thresholds of the tools.
    }
    \label{fig:llm_detectors}
\end{figure}

Considering the relatively short length of individual tweets from the fox8 bots, we concatenate them per user and run the consolidated text through OpenAI's detector.
We then plot the distribution of the OpenAI detector scores in Figure~\ref{fig:llm_detectors}(a).
Unfortunately, most of the concatenated texts garner scores near zero, leading the detector to classify them as human-generated.
Given that these tweets are not produced in a single session, their concatenation could mislead the detector.
Consequently, we explore methodologies for tweet-level detection.

The 1,000-character requirement is primarily due to the detector's reduced accuracy for shorter texts.
It is only reinforced on the webpage.
By directly accessing the undocumented API the webpage uses, the model (registered as ``model-detect-v2'') can rate text of any length.
Hence, we input each tweet from the fox8 bots into the detector and show the score distribution in Figure~\ref{fig:llm_detectors}(a).
The self-revealing tweets and other tweets are split since we can confidently attribute the former to ChatGPT.
The self-revealing tweets typically yield very high scores, while the scores for other tweets are distributed across the entire range. 

Next, we examine the performance of GPTZero.
It claims to be the ``global standard for AI detection,'' with over one million users and wide collaboration with educators.
According to its FAQ,\footnote{gptzero.me/faq (Accessed May 2023)} GPTZero is a ``classification model that predicts whether a document was written by a large language model, providing predictions on a sentence, paragraph, and document level.''
This suggests that it is a black-box detection method as well.

We use its API to analyze the tweets of the fox8 bots.
Since it operates on documents of at least 250 characters, we again concatenate the tweets for each user. 
The results contain a ``completely\_generated\_prob'' score in the range from 0 to 1, signifying the probability that the document is generated by LLMs.
Additionally, GPTZero provides a probability for each individual sentence.
The official documentation suggests using 0.65 as the threshold to dichotomize the probabilities.
We show the distributions of overall and sentence-level probabilities for the self-revealing and other tweets in Figure~\ref{fig:llm_detectors}(b).
All probabilities are near zero.

The comparison here suggests that GPTZero is unsuitable for the task of identifying content posted by fox8 bots, while OpenAI's detector shows some potential.
Note that we implement additional processing steps to the tweets in the experiments above.
We exclude retweets since they originate from other accounts.
For the rest of the tweets (i.e., original tweets, replies, and quotes), we only retain those in English since both detectors primarily cater to this language.
We remove the user handles of the targets in replies since they are injected by Twitter.
The links in the tweets are also removed.

\subsection{Detecting LLM-powered bots}

Due to the valuable indicators provided by OpenAI's AI text detector at the tweet level, let us explore the feasibility of building a tool to detect LLM-powered bots on top of it.
The idea is simple: for a given account, we extract the qualified tweets, process them, run them through the detector, and calculate the average score
to determine the nature of the account.

\begin{figure}
    \centering
    \includegraphics[width=\columnwidth]{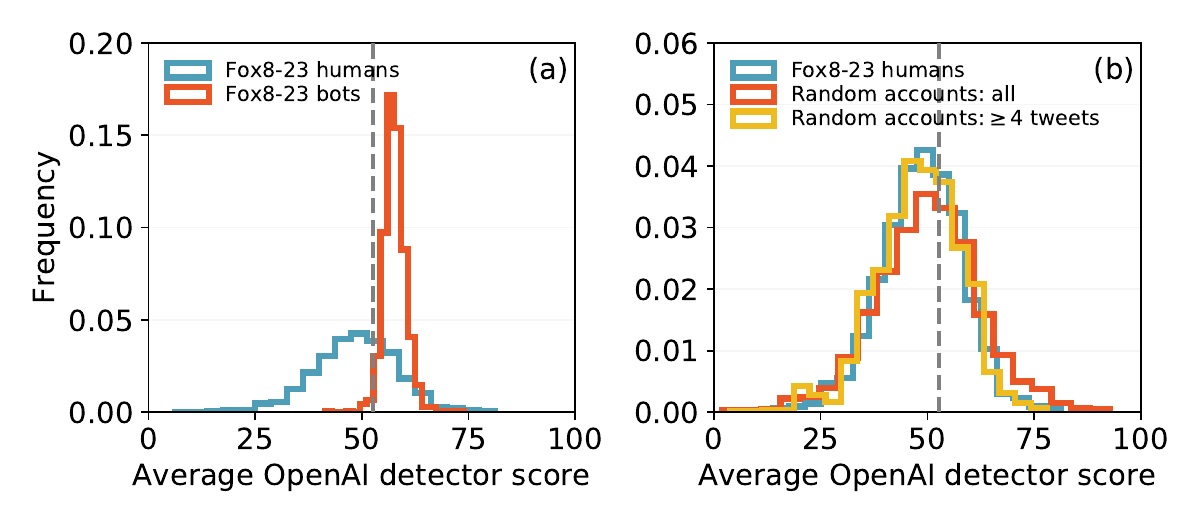}
    \caption{
    Using OpenAI's AI text detector to identify fox8 bots.
    (a) Distributions of the average OpenAI detector score for human and bot accounts in the \texttt{fox8-23} dataset.
    (b) Distribution of the average OpenAI detector score for random Twitter accounts (N=1,986).
    For comparison, we also show the distribution for the subset of random accounts with at least four qualified tweets (N=1,269) and the results for the human accounts in \texttt{fox8-23}.
    The vertical lines mark the optimal threshold for distinguishing between the humans and bots in \texttt{fox8-23} (see main text for details).
    }
    \label{fig:openai_detector_performance}
\end{figure}

We use the full \texttt{fox8-23} dataset in this experiment.
The distribution of the final scores at the account level can be found in Figure~\ref{fig:openai_detector_performance}(a).
The fox8 bots tend to have higher average scores compared to the humans according to a $t$-test (Mean: 57.7 vs. 48.6, $t=30.6, p<0.001$), but human users exhibit a larger standard deviation (SD: 2.6 vs.~9.7).
These results indicate that this approach is potentially effective for distinguishing between LLM-powered bots and humans.
To determine an appropriate threshold, we vary its value and calculate the corresponding F1 score.
When the threshold is set to 52.7, the F1 score is maximized, reaching  0.84.

Let us apply this method in the field to test its effectiveness further.
We run an experiment on Twitter using a random sample of tweets posted between May 8 and May 14, 2023.
We identify the unique accounts and sample 4,000 for further analysis.
Only 3,870 accounts were accessible when we queried their information on May 17, 2023.
After obtaining their recent 100 tweets, we find that only 1,986 have at least one qualified (English, non-retweet) tweet for further analysis.

We calculate the average OpenAI detector scores for these 1,986 accounts and show their distribution in Figure~\ref{fig:openai_detector_performance}(b).
For reference, we also show the results for the human accounts in \texttt{fox8-23}.
We find that the random accounts have slightly higher scores than the \texttt{fox8-23} humans according to a $t$-test ($t=3.4, p<0.001$).
Applying the threshold from earlier (52.7)  leads to 815 of the random accounts being labeled as bots.
But these might include many false positives.
The \texttt{fox8-23} dataset contains the same amount of bots and humans, however, we believe LLM-powered bots are still rare on Twitter as of today.
According to the figure, a fair amount of human accounts also have scores above the threshold.

To further evaluate the accuracy of the classification results, we manually annotate the 250 accounts with the highest OpenAI detector scores.
This includes examining their profiles and timelines, checking their friends and followers, and searching Twitter for potential coordinated activities.
Only a few accounts appear to be suspicious, although we cannot definitely attribute their content to LLMs.

Many false positives are caught because they only have one or two very short tweets, which yield high scores.
For example, both the terms ``thank you'' and ``amen'' have scored over 60.
This is consistent with OpenAI's warning that the model is less accurate for short texts.
These accounts also increase the average score of random accounts.
To illustrate, we show the results for a subset of the random accounts with at least four qualified tweets in Figure~\ref{fig:openai_detector_performance}(b), and their scores are not significantly different from those of the \texttt{fox8-23} humans ($t=-1.4, p=0.15$).

\section{Conclusion and discussion}

This paper presents a case study about a Twitter botnet that employs ChatGPT for content generation.
Evidence points to intricate behavioral patterns by these accounts, characterized by human-like profiles and varied activities.
Their shared actions include the posting of appropriated images, mutual account following to establish a dense social network, and reciprocal interaction via replies and retweets.
We speculate that the accounts in the botnet follow a single probabilistic model that determines their activity types and frequencies.
ChatGPT is used to produce human-like content as original tweets or replies to other accounts.
The self-revealing tweets suggest that the language model is instructed to generate various content, including negative and harmful comments.
Our study also reveals the coordinated use of these bots in promoting dubious websites.

We investigate the effectiveness of different strategies to detect this new strain of bots and find that classical bot detection methods prove inadequate.
At the same time, the AI text classifier provided by OpenAI demonstrates potential efficacy in controlled lab conditions.
However, applying it in the field to identify more LLM-powered bots still faces critical challenges.
First, it is unreliable for non-English content~\citep{liang2023gpt} and short texts, considerably narrowing the scope of accounts it can process.
Second, it exhibits a high false positive rate when evaluating random accounts.
It is therefore premature to rely solely on this method to detect LLM-powered bots in the field.

Our analysis has some limitations.
Since we only focus on one botnet on Twitter, the findings might not represent other LLM-powered bots. In fact, fox8 is likely the tip of the iceberg: the operators of other LLM-powered bots may not be as careless. 
Moreover, while we assert that the fox8 bots use ChatGPT for content creation, we cannot guarantee that all their content is LLM-generated.
Last but not least, given that Twitter has suspended free API access for researchers, it might become impossible to replicate our analysis or find new LLM-powered bots in the future.

Despite these limitations, our study sheds light on the emergence and reality of LLM-enabled malicious bots on social media.
Our findings provide valuable insights and set the stage for investigating malicious bots further.
Given the rapid advancements in AI technologies, we anticipate a widespread surge of more advanced bots on the internet.
Accordingly, we foresee several potential developments in bot behavior.

Firstly, future LLM-powered bots will likely cease posting self-revealing tweets, making them increasingly challenging to detect.
Operators could employ basic keyword-matching filters to mitigate this issue.
Furthermore, the swift advancements in open-source LLMs could incentivize operators to utilize models lacking safeguards, or even train models specifically for malicious purposes~\citep{jin2023darkbert}.

Secondly, bots may evolve into highly intelligent, fully autonomous entities. 
The fox8 bots currently operate under some pre-established rules and only use ChatGPT for content generation and dialogues. 
However, emerging research indicates that LLMs can facilitate the development of autonomous agents capable of independently processing exposed information, making autonomous decisions, and utilizing tools such as APIs and search engines~\citep{li2023camel,park2023generative,wang2023voyager}.
Open-source implementations, such as AutoGPT\footnote{github.com/Significant-Gravitas/Auto-GPT} and BabyAGI,\footnote{github.com/yoheinakajima/babyagi} make it straightforward to integrate these agents with Twitter accounts.

Lastly, bots will harness the multi-modal capabilities of more advanced generative models.
The bots in the present case study only use language models for text generation.
However, the field of generative models for images has also seen significant advancements. 
For instance,  Generative Adversarial Networks can already create realistic human faces~\citep{karras2019style,karras2020analyzing} that humans fail to identify~\citep{nightingale2022ai}, while the stable diffusion algorithms are capable of producing a variety of images~\citep{rombach2022high}.
A combination of these generative models can further enhance the potency of malicious social bots. 

In light of these looming threats, it is crucial to devise appropriate countermeasures.
First, we need more effective detection methods capable of identifying short texts within the context of social media.
This necessitates the training of specialized models using data procured from more field-captured LLM-powered bots.
To this end, one could search for other phrases, such as ``I'm sorry, I cannot generate,'' that LLMs use when refusing to comply with the prompts~\citep{dekens2023practical} on other social media or websites.\footnote{nytimes.com/2023/05/19/technology/ai-generated-content-discovered-on-news-sites-content-farms-and-product-reviews.html (Accessed May 2023)}
Once we gain a deeper understanding of the instructions fed to these bots, we can also utilize LLMs to generate additional texts independently.
Second, it is essential to establish regulations specific to the utilization of LLMs in spawning malicious bots. For example, a platform might be required to challenge an account to prove a piece of content is organic before it becomes visible to a large audience~\citep{Menczer2023AI-harms}. 
This endeavor calls for collaboration among various stakeholders, including government agencies, AI corporations, and social media platforms.
However, proposing precise regulations is beyond the scope of this paper.
Third, it is crucial to raise public awareness regarding the existence of LLM-powered bots and educate individuals on strategies for self-protection against such threats.
Nevertheless, this initiative should be carried out carefully to avoid unintended consequences as recent research indicates that preemptively informing users about the existence of social bots may amplify their existing cognitive biases~\citep{yan2022exposure}.

\subsection*{Acknowledgements}

We thank Twitter users @conspirator0 and @jsrailton for the inspiration of searching ``as an ai language model'' to identify LLM-powered bots on Twitter. 
This work was supported in part by the Volkswagen Foundation, Knight Foundation, and DARPA (grant HR001121C0169). 

\bibliography{main}

\begin{thebibliography}{76}
\providecommand{\natexlab}[1]{#1}
\providecommand{\url}[1]{\texttt{#1}}
\expandafter\ifx\csname urlstyle\endcsname\relax
  \providecommand{\doi}[1]{doi: #1}\else
  \providecommand{\doi}{doi: \begingroup \urlstyle{rm}\Url}\fi

\bibitem[Assenmacher et~al.(2020)Assenmacher, Clever, Frischlich, Quandt,
  Trautmann, and Grimme]{assenmacher2020demystifying}
Dennis Assenmacher, Lena Clever, Lena Frischlich, Thorsten Quandt, Heike
  Trautmann, and Christian Grimme.
\newblock Demystifying social bots: On the intelligence of automated social
  media actors.
\newblock \emph{Social Media+ Society}, 6\penalty0 (3):\penalty0
  2056305120939264, 2020.

\bibitem[Ayers et~al.(2023)Ayers, Poliak, Dredze, Leas, Zhu, Kelley, Faix,
  Goodman, Longhurst, Hogarth, et~al.]{ayers2023comparing}
John~W Ayers, Adam Poliak, Mark Dredze, Eric~C Leas, Zechariah Zhu, Jessica~B
  Kelley, Dennis~J Faix, Aaron~M Goodman, Christopher~A Longhurst, Michael
  Hogarth, et~al.
\newblock Comparing physician and artificial intelligence chatbot responses to
  patient questions posted to a public social media forum.
\newblock \emph{JAMA Internal Medicine}, 2023.

\bibitem[Bahrini et~al.(2023)Bahrini, Khamoshifar, Abbasimehr, Riggs, Esmaeili,
  Majdabadkohne, and Pasehvar]{bahrini2023chatgpt}
Aram Bahrini, Mohammadsadra Khamoshifar, Hossein Abbasimehr, Robert~J Riggs,
  Maryam Esmaeili, Rastin~Mastali Majdabadkohne, and Morteza Pasehvar.
\newblock Chatgpt: Applications, opportunities, and threats.
\newblock \emph{Preprint arXiv:2304.09103}, 2023.

\bibitem[Bommasani et~al.(2021)Bommasani, Hudson, Adeli, Altman, Arora, von
  Arx, Bernstein, Bohg, Bosselut, Brunskill,
  et~al.]{bommasani2021opportunities}
Rishi Bommasani, Drew~A Hudson, Ehsan Adeli, Russ Altman, Simran Arora, Sydney
  von Arx, Michael~S Bernstein, Jeannette Bohg, Antoine Bosselut, Emma
  Brunskill, et~al.
\newblock On the opportunities and risks of foundation models.
\newblock \emph{Preprint arXiv:2108.07258}, 2021.

\bibitem[Brundage et~al.(2018)Brundage, Avin, Clark, Toner, Eckersley,
  Garfinkel, Dafoe, Scharre, Zeitzoff, Filar, et~al.]{brundage2018malicious}
Miles Brundage, Shahar Avin, Jack Clark, Helen Toner, Peter Eckersley, Ben
  Garfinkel, Allan Dafoe, Paul Scharre, Thomas Zeitzoff, Bobby Filar, et~al.
\newblock The malicious use of artificial intelligence: Forecasting,
  prevention, and mitigation.
\newblock \emph{Preprint arXiv:1802.07228}, 2018.

\bibitem[Buchanan et~al.(2021)Buchanan, Lohn, and Musser]{buchanan2021truth}
Ben Buchanan, Andrew Lohn, and Micah Musser.
\newblock \emph{Truth, lies, and automation: How language models could change
  disinformation}.
\newblock Center for Security and Emerging Technology, 2021.

\bibitem[Chakraborty et~al.(2023)Chakraborty, Bedi, Zhu, An, Manocha, and
  Huang]{chakraborty2023possibilities}
Souradip Chakraborty, Amrit~Singh Bedi, Sicheng Zhu, Bang An, Dinesh Manocha,
  and Furong Huang.
\newblock On the possibilities of ai-generated text detection.
\newblock \emph{Preprint arXiv:2304.04736}, 2023.

\bibitem[Chavoshi et~al.(2016)Chavoshi, Hamooni, and Mueen]{chavoshi2016debot}
Nikan Chavoshi, Hossein Hamooni, and Abdullah Mueen.
\newblock Debot: Twitter bot detection via warped correlation.
\newblock In \emph{IEEE International Conference on Data Mining (ICDM)}, pages
  817--822, 2016.

\bibitem[Clark et~al.(2021)Clark, August, Serrano, Haduong, Gururangan, and
  Smith]{clark2021all}
Elizabeth Clark, Tal August, Sofia Serrano, Nikita Haduong, Suchin Gururangan,
  and Noah~A Smith.
\newblock All that's `human'is not gold: Evaluating human evaluation of
  generated text.
\newblock \emph{Preprint arXiv:2107.00061}, 2021.

\bibitem[Conover et~al.(2012)Conover, Gon\c{c}alves, Flammini, and
  Menczer]{conover12partisan}
Michael~D Conover, Bruno Gon\c{c}alves, Alessandro Flammini, and Filippo
  Menczer.
\newblock Partisan asymmetries in online political activity.
\newblock \emph{EPJ Data Science}, 1:\penalty0 6, 2012.
\newblock \doi{10.1140/epjds6}.
\newblock URL \url{http://dx.doi.org/10.1140/epjds6}.

\bibitem[Cresci(2020)]{cresci2020decade}
Stefano Cresci.
\newblock A decade of social bot detection.
\newblock \emph{Communications of the ACM}, 63\penalty0 (10):\penalty0 72--83,
  2020.

\bibitem[Crothers et~al.(2022)Crothers, Japkowicz, and
  Viktor]{crothers2022machine}
Evan Crothers, Nathalie Japkowicz, and Herna Viktor.
\newblock Machine generated text: A comprehensive survey of threat models and
  detection methods.
\newblock \emph{Preprint arXiv:2210.07321}, 2022.

\bibitem[Dekens(2023)]{dekens2023practical}
Nico Dekens.
\newblock A practical guide for osint investigators to combat disinformation
  and fake reviews driven by ai (chatgpt).
\newblock
  \url{https://info.shadowdragon.io/hubfs/SD_APracticalGuide_WhitePaper-1.pdf},
  2023.
\newblock (Accessed May 2023).

\bibitem[Fagni et~al.(2021)Fagni, Falchi, Gambini, Martella, and
  Tesconi]{fagni2021tweepfake}
Tiziano Fagni, Fabrizio Falchi, Margherita Gambini, Antonio Martella, and
  Maurizio Tesconi.
\newblock Tweepfake: About detecting deepfake tweets.
\newblock \emph{PLOS ONE}, 16\penalty0 (5):\penalty0 1--16, 05 2021.
\newblock \doi{10.1371/journal.pone.0251415}.
\newblock URL \url{https://doi.org/10.1371/journal.pone.0251415}.

\bibitem[Ferrara(2023)]{ferrara2023social}
Emilio Ferrara.
\newblock Social bot detection in the age of chatgpt: Challenges and
  opportunities.
\newblock \emph{First Monday}, 2023.

\bibitem[Ferrara et~al.(2016)Ferrara, Varol, Davis, Menczer, and
  Flammini]{ferrara2016rise}
Emilio Ferrara, Onur Varol, Clayton Davis, Filippo Menczer, and Alessandro
  Flammini.
\newblock The rise of social bots.
\newblock \emph{Communications of the ACM}, 59\penalty0 (7):\penalty0 96--104,
  2016.

\bibitem[Ferrara et~al.(2020)Ferrara, Chang, Chen, Muric, and
  Patel]{ferrara2020characterizing}
Emilio Ferrara, Herbert Chang, Emily Chen, Goran Muric, and Jaimin Patel.
\newblock Characterizing social media manipulation in the 2020 us presidential
  election.
\newblock \emph{First Monday}, 2020.

\bibitem[Fr{\"o}hling and Zubiaga(2021)]{frohling2021feature}
Leon Fr{\"o}hling and Arkaitz Zubiaga.
\newblock Feature-based detection of automated language models: tackling gpt-2,
  gpt-3 and grover.
\newblock \emph{PeerJ Computer Science}, 7:\penalty0 e443, 2021.

\bibitem[Gambini et~al.(2022)Gambini, Fagni, Falchi, and
  Tesconi]{gambini2022pushing}
Margherita Gambini, Tiziano Fagni, Fabrizio Falchi, and Maurizio Tesconi.
\newblock On pushing deepfake tweet detection capabilities to the limits.
\newblock In \emph{14th ACM Web Science Conference 2022}, pages 154--163, 2022.

\bibitem[Gehrmann et~al.(2019)Gehrmann, Strobelt, and Rush]{gehrmann2019gltr}
Sebastian Gehrmann, Hendrik Strobelt, and Alexander~M Rush.
\newblock Gltr: Statistical detection and visualization of generated text.
\newblock \emph{Preprint arXiv:1906.04043}, 2019.

\bibitem[Giglietto et~al.(2020)Giglietto, Righetti, Rossi, and
  Marino]{giglietto2020coordinated}
Fabio Giglietto, Nicola Righetti, Luca Rossi, and Giada Marino.
\newblock Coordinated link sharing behavior as a signal to surface sources of
  problematic information on facebook.
\newblock In \emph{International Conference on Social Media and Society}, pages
  85--91, 2020.

\bibitem[Gilani et~al.(2017)Gilani, Farahbakhsh, Tyson, Wang, and
  Crowcroft]{gilani2017bots}
Zafar Gilani, Reza Farahbakhsh, Gareth Tyson, Liang Wang, and Jon Crowcroft.
\newblock {Of bots and humans (on Twitter}).
\newblock In \emph{{Proceedings of the International Conference on Advances in
  Social Networks Analysis and Mining}}, pages 349--354. ACM, 2017.

\bibitem[Goldstein et~al.(2023)Goldstein, Sastry, Musser, DiResta, Gentzel, and
  Sedova]{goldstein2023generative}
Josh~A Goldstein, Girish Sastry, Micah Musser, Renee DiResta, Matthew Gentzel,
  and Katerina Sedova.
\newblock Generative language models and automated influence operations:
  Emerging threats and potential mitigations.
\newblock \emph{Preprint arXiv:2301.04246}, 2023.

\bibitem[Grimme et~al.(2022)Grimme, Pohl, Cresci, L{\"u}ling, and
  Preuss]{grimme2022new}
Christian Grimme, Janina Pohl, Stefano Cresci, Ralf L{\"u}ling, and Mike
  Preuss.
\newblock New automation for social bots: From trivial behavior to ai-powered
  communication.
\newblock In Francesca Spezzano, Adriana Amaral, Davide Ceolin, Lisa Fazio, and
  Edoardo Serra, editors, \emph{Disinformation in Open Online Media}, pages
  79--99, Cham, 2022. Springer International Publishing.
\newblock ISBN 978-3-031-18253-2.

\bibitem[Guembe et~al.(2022)Guembe, Azeta, Misra, Osamor, Fernandez-Sanz, and
  Pospelova]{guembe2022emerging}
Blessing Guembe, Ambrose Azeta, Sanjay Misra, Victor~Chukwudi Osamor, Luis
  Fernandez-Sanz, and Vera Pospelova.
\newblock The emerging threat of ai-driven cyber attacks: A review.
\newblock \emph{Applied Artificial Intelligence}, 36\penalty0 (1):\penalty0
  2037254, 2022.

\bibitem[Guo et~al.(2023)Guo, Zhang, Wang, Jiang, Nie, Ding, Yue, and
  Wu]{guo2023close}
Biyang Guo, Xin Zhang, Ziyuan Wang, Minqi Jiang, Jinran Nie, Yuxuan Ding,
  Jianwei Yue, and Yupeng Wu.
\newblock How close is chatgpt to human experts? comparison corpus, evaluation,
  and detection.
\newblock \emph{Preprint arXiv:2301.07597}, 2023.

\bibitem[Hanley and Durumeric(2023)]{hanley2023machine}
Hans~WA Hanley and Zakir Durumeric.
\newblock Machine-made media: Monitoring the mobilization of machine-generated
  articles on misinformation and mainstream news websites.
\newblock \emph{Preprint arXiv:2305.09820}, 2023.

\bibitem[Hazell(2023)]{hazell2023large}
Julian Hazell.
\newblock Large language models can be used to effectively scale spear phishing
  campaigns.
\newblock \emph{Preprint arXiv:2305.06972}, 2023.

\bibitem[Heidari and Jones(2020)]{heidari2020using}
Maryam Heidari and James~H Jones.
\newblock Using bert to extract topic-independent sentiment features for social
  media bot detection.
\newblock In \emph{IEEE Annual Ubiquitous Computing, Electronics \& Mobile
  Communication Conference (UEMCON)}, pages 0542--0547. IEEE, 2020.

\bibitem[Jakesch et~al.(2023)Jakesch, Hancock, and Naaman]{jakesch2023human}
Maurice Jakesch, Jeffrey~T. Hancock, and Mor Naaman.
\newblock Human heuristics for ai-generated language are flawed.
\newblock \emph{Proceedings of the National Academy of Sciences}, 120\penalty0
  (11):\penalty0 e2208839120, 2023.

\bibitem[Jamison et~al.(2019)Jamison, Broniatowski, and
  Quinn]{jamison_malicious_2019}
Amelia~M. Jamison, David~A. Broniatowski, and Sandra~Crouse Quinn.
\newblock Malicious {Actors} on {Twitter}: {A} {Guide} for {Public} {Health}
  {Researchers}.
\newblock \emph{American Journal of Public Health}, 109\penalty0 (5):\penalty0
  688--692, 2019.
\newblock ISSN 0090-0036.

\bibitem[Jin et~al.(2023)Jin, Jang, Cui, Chung, Lee, and Shin]{jin2023darkbert}
Youngjin Jin, Eugene Jang, Jian Cui, Jin-Woo Chung, Yongjae Lee, and Seungwon
  Shin.
\newblock Darkbert: A language model for the dark side of the internet.
\newblock \emph{Preprint arXiv:2305.08596}, 2023.

\bibitem[Karras et~al.(2019)Karras, Laine, and Aila]{karras2019style}
Tero Karras, Samuli Laine, and Timo Aila.
\newblock A style-based generator architecture for generative adversarial
  networks.
\newblock In \emph{Proceedings of the IEEE/CVF Conference on Computer Vision
  and Pattern Recognition}, pages 4401--4410, 2019.

\bibitem[Karras et~al.(2020)Karras, Laine, Aittala, Hellsten, Lehtinen, and
  Aila]{karras2020analyzing}
Tero Karras, Samuli Laine, Miika Aittala, Janne Hellsten, Jaakko Lehtinen, and
  Timo Aila.
\newblock Analyzing and improving the image quality of stylegan.
\newblock In \emph{Proceedings of the IEEE/CVF Conference on Computer Vision
  and Pattern Recognition}, pages 8110--8119, 2020.

\bibitem[Kasneci et~al.(2023)Kasneci, Se{\ss}ler, K{\"u}chemann, Bannert,
  Dementieva, Fischer, Gasser, Groh, G{\"u}nnemann, H{\"u}llermeier,
  et~al.]{learning2023kasneci}
Enkelejda Kasneci, Kathrin Se{\ss}ler, Stefan K{\"u}chemann, Maria Bannert,
  Daryna Dementieva, Frank Fischer, Urs Gasser, Georg Groh, Stephan
  G{\"u}nnemann, Eyke H{\"u}llermeier, et~al.
\newblock Chatgpt for good? on opportunities and challenges of large language
  models for education.
\newblock \emph{Learning and Individual Differences}, 103:\penalty0 102274,
  2023.
\newblock ISSN 1041-6080.

\bibitem[Keller et~al.(2017)Keller, Schoch, Stier, and
  Yang]{keller2017manipulate}
Franziska Keller, David Schoch, Sebastian Stier, and JungHwan Yang.
\newblock How to manipulate social media: Analyzing political astroturfing
  using ground truth data from south korea.
\newblock In \emph{Proceedings of the International AAAI Conference on Web and
  Social Media}, volume~11, 2017.

\bibitem[Keller et~al.(2020)Keller, Schoch, Stier, and
  Yang]{keller2020political}
Franziska~B Keller, David Schoch, Sebastian Stier, and JungHwan Yang.
\newblock Political astroturfing on twitter: How to coordinate a disinformation
  campaign.
\newblock \emph{Political Communication}, 37\penalty0 (2):\penalty0 256--280,
  2020.

\bibitem[Kirchenbauer et~al.(2023)Kirchenbauer, Geiping, Wen, Katz, Miers, and
  Goldstein]{kirchenbauer2023watermark}
John Kirchenbauer, Jonas Geiping, Yuxin Wen, Jonathan Katz, Ian Miers, and Tom
  Goldstein.
\newblock A watermark for large language models.
\newblock \emph{Preprint arXiv:2301.10226}, 2023.

\bibitem[Kreps et~al.(2022)Kreps, McCain, and Brundage]{kreps2022news}
Sarah Kreps, R.~Miles McCain, and Miles Brundage.
\newblock All the news that’s fit to fabricate: Ai-generated text as a tool
  of media misinformation.
\newblock \emph{Journal of Experimental Political Science}, 9\penalty0
  (1):\penalty0 104--117, 2022.
\newblock \doi{10.1017/XPS.2020.37}.

\bibitem[Krishna et~al.(2023)Krishna, Song, Karpinska, Wieting, and
  Iyyer]{krishna2023paraphrasing}
Kalpesh Krishna, Yixiao Song, Marzena Karpinska, John Wieting, and Mohit Iyyer.
\newblock Paraphrasing evades detectors of ai-generated text, but retrieval is
  an effective defense.
\newblock \emph{Preprint arXiv:2303.13408}, 2023.

\bibitem[Kucharavy et~al.(2023)Kucharavy, Schillaci, Mar{\'e}chal, W{\"u}rsch,
  Dolamic, Sabonnadiere, David, Mermoud, and
  Lenders]{kucharavy2023fundamentals}
Andrei Kucharavy, Zachary Schillaci, Lo{\"\i}c Mar{\'e}chal, Maxime W{\"u}rsch,
  Ljiljana Dolamic, Remi Sabonnadiere, Dimitri~Percia David, Alain Mermoud, and
  Vincent Lenders.
\newblock Fundamentals of generative large language models and perspectives in
  cyber-defense.
\newblock \emph{Preprint arXiv:2303.12132}, 2023.

\bibitem[Kudugunta and Ferrara(2018)]{kudugunta2018deep}
Sneha Kudugunta and Emilio Ferrara.
\newblock Deep neural networks for bot detection.
\newblock \emph{Information Sciences}, 467:\penalty0 312--322, 2018.

\bibitem[Kumarage et~al.(2023)Kumarage, Garland, Bhattacharjee, Trapeznikov,
  Ruston, and Liu]{kumarage2023stylometric}
Tharindu Kumarage, Joshua Garland, Amrita Bhattacharjee, Kirill Trapeznikov,
  Scott Ruston, and Huan Liu.
\newblock Stylometric detection of ai-generated text in twitter timelines.
\newblock \emph{Preprint arXiv:2303.03697}, 2023.

\bibitem[Li et~al.(2023)Li, Hammoud, Itani, Khizbullin, and
  Ghanem]{li2023camel}
Guohao Li, Hasan Abed Al~Kader Hammoud, Hani Itani, Dmitrii Khizbullin, and
  Bernard Ghanem.
\newblock Camel: Communicative agents for" mind" exploration of large scale
  language model society.
\newblock \emph{Preprint arXiv:2303.17760}, 2023.

\bibitem[Liang et~al.(2023)Liang, Yuksekgonul, Mao, Wu, and Zou]{liang2023gpt}
Weixin Liang, Mert Yuksekgonul, Yining Mao, Eric Wu, and James Zou.
\newblock Gpt detectors are biased against non-native english writers.
\newblock \emph{Preprint arXiv:2304.02819}, 2023.

\bibitem[Liu et~al.(2023)Liu, Yuan, Fu, Jiang, Hayashi, and Neubig]{liu2023pre}
Pengfei Liu, Weizhe Yuan, Jinlan Fu, Zhengbao Jiang, Hiroaki Hayashi, and
  Graham Neubig.
\newblock Pre-train, prompt, and predict: A systematic survey of prompting
  methods in natural language processing.
\newblock \emph{ACM Computing Surveys}, 55\penalty0 (9):\penalty0 1--35, 2023.

\bibitem[Marlow et~al.(2020)Marlow, Miller, and Roberts]{marlow_twitter_2020}
Thomas Marlow, Sean Miller, and J.~Timmons Roberts.
\newblock Twitter {Discourses} on {Climate} {Change}: {Exploring} {Topics} and
  the {Presence} of {Bots}.
\newblock \emph{SocArXiv}, 2020.
\newblock \doi{10.31235/osf.io/h6ktm}.

\bibitem[Menczer et~al.(2023)Menczer, Crandall, Ahn, and
  Kapadia]{Menczer2023AI-harms}
Filippo Menczer, David Crandall, Yong-Yeol Ahn, and Apu Kapadia.
\newblock {Addressing the harms of AI-generated inauthentic content}.
\newblock \emph{Nature Machine Intelligence}, 5:\penalty0 679--680, 2023.
\newblock \doi{10.1038/s42256-023-00690-w}.
\newblock URL \url{https://rdcu.be/dgGfk}.

\bibitem[Mitchell et~al.(2023)Mitchell, Lee, Khazatsky, Manning, and
  Finn]{mitchell2023detectgpt}
Eric Mitchell, Yoonho Lee, Alexander Khazatsky, Christopher~D Manning, and
  Chelsea Finn.
\newblock Detectgpt: Zero-shot machine-generated text detection using
  probability curvature.
\newblock \emph{Preprint arXiv:2301.11305}, 2023.

\bibitem[Nightingale and Farid(2022)]{nightingale2022ai}
Sophie~J Nightingale and Hany Farid.
\newblock Ai-synthesized faces are indistinguishable from real faces and more
  trustworthy.
\newblock \emph{Proceedings of the National Academy of Sciences}, 119\penalty0
  (8):\penalty0 e2120481119, 2022.

\bibitem[Nizzoli et~al.(2021)Nizzoli, Tardelli, Avvenuti, Cresci, and
  Tesconi]{nizzoli2021coordinated}
Leonardo Nizzoli, Serena Tardelli, Marco Avvenuti, Stefano Cresci, and Maurizio
  Tesconi.
\newblock Coordinated behavior on social media in 2019 {UK} general election.
\newblock In \emph{Proceedings of the International AAAI Conference on Web and
  Social Media}, volume~15, pages 443--454, 2021.

\bibitem[OpenAI(2023)]{AITextClassifier}
OpenAI.
\newblock Ai text classifier.
\newblock https://beta.openai.com/ai-text-classifier, January 2023.
\newblock URL \url{https://beta.openai.com/ai-text-classifier}.

\bibitem[Ouyang et~al.(2022)Ouyang, Wu, Jiang, Almeida, Wainwright, Mishkin,
  Zhang, Agarwal, Slama, Ray, et~al.]{ouyang2022training}
Long Ouyang, Jeffrey Wu, Xu~Jiang, Diogo Almeida, Carroll Wainwright, Pamela
  Mishkin, Chong Zhang, Sandhini Agarwal, Katarina Slama, Alex Ray, et~al.
\newblock Training language models to follow instructions with human feedback.
\newblock \emph{Advances in Neural Information Processing Systems},
  35:\penalty0 27730--27744, 2022.

\bibitem[Pacheco et~al.(2021)Pacheco, Hui, Torres-Lugo, Truong, Flammini, and
  Menczer]{pacheco2020uncovering}
Diogo Pacheco, Pik-Mai Hui, Christopher Torres-Lugo, Bao~Tran Truong,
  Alessandro Flammini, and Filippo Menczer.
\newblock Uncovering coordinated networks on social media: Methods and case
  studies.
\newblock In \emph{Proceedings of the International AAAI Conference on Web and
  Social Media}, volume~15, pages 455--466, 2021.

\bibitem[Park et~al.(2023)Park, O'Brien, Cai, Morris, Liang, and
  Bernstein]{park2023generative}
Joon~Sung Park, Joseph~C O'Brien, Carrie~J Cai, Meredith~Ringel Morris, Percy
  Liang, and Michael~S Bernstein.
\newblock Generative agents: Interactive simulacra of human behavior.
\newblock \emph{Preprint arXiv:2304.03442}, 2023.

\bibitem[Qin et~al.(2023)Qin, Zhang, Zhang, Chen, Yasunaga, and
  Yang]{qin2023chatgpt}
Chengwei Qin, Aston Zhang, Zhuosheng Zhang, Jiaao Chen, Michihiro Yasunaga, and
  Diyi Yang.
\newblock {Is ChatGPT a general-purpose natural language processing task
  solver?}
\newblock \emph{Preprint arXiv:2302.06476}, 2023.

\bibitem[Rombach et~al.(2022)Rombach, Blattmann, Lorenz, Esser, and
  Ommer]{rombach2022high}
Robin Rombach, Andreas Blattmann, Dominik Lorenz, Patrick Esser, and Bj{\"o}rn
  Ommer.
\newblock High-resolution image synthesis with latent diffusion models.
\newblock In \emph{Proceedings of the IEEE/CVF Conference on Computer Vision
  and Pattern Recognition}, pages 10684--10695, 2022.

\bibitem[Sadasivan et~al.(2023)Sadasivan, Kumar, Balasubramanian, Wang, and
  Feizi]{sadasivan2023can}
Vinu~Sankar Sadasivan, Aounon Kumar, Sriram Balasubramanian, Wenxiao Wang, and
  Soheil Feizi.
\newblock Can ai-generated text be reliably detected?
\newblock \emph{Preprint arXiv:2303.11156}, 2023.

\bibitem[Saravani et~al.(2021)Saravani, Ray, and Ray]{saravani2021automated}
Sina~Mahdipour Saravani, Indrajit Ray, and Indrakshi Ray.
\newblock Automated identification of social media bots using deepfake text
  detection.
\newblock In \emph{Information Systems Security: International Conference
  (ICISS)}, pages 111--123. Springer, 2021.

\bibitem[Sayyadiharikandeh et~al.(2020)Sayyadiharikandeh, Varol, Yang,
  Flammini, and Menczer]{botometerv4-2020}
Mohsen Sayyadiharikandeh, Onur Varol, Kai-Cheng Yang, Alessandro Flammini, and
  Filippo Menczer.
\newblock Detection of novel social bots by ensembles of specialized
  classifiers.
\newblock In \emph{Proc. 29th ACM International Conference on Information \&
  Knowledge Management (CIKM)}, pages 2725--2732, 2020.
\newblock \doi{10.1145/3340531.3412698}.
\newblock URL \url{https://doi.org/10.1145/3340531.3412698}.

\bibitem[Shao et~al.(2018)Shao, Ciampaglia, Varol, Yang, Flammini, and
  Menczer]{shao2018spread}
Chengcheng Shao, Giovanni~Luca Ciampaglia, Onur Varol, Kai-Cheng Yang,
  Alessandro Flammini, and Filippo Menczer.
\newblock The spread of low-credibility content by social bots.
\newblock \emph{Nature communications}, 9\penalty0 (1):\penalty0 4787, 2018.

\bibitem[Spitale et~al.(2023)Spitale, Biller-Andorno, and
  Germani]{spitale2023ai}
Giovanni Spitale, Nikola Biller-Andorno, and Federico Germani.
\newblock Ai model gpt-3 (dis) informs us better than humans.
\newblock \emph{Preprint arXiv:2301.11924}, 2023.

\bibitem[Tang et~al.(2023)Tang, Chuang, and Hu]{tang2023science}
Ruixiang Tang, Yu-Neng Chuang, and Xia Hu.
\newblock {The science of detecting LLM-generated texts}.
\newblock \emph{Preprint arXiv:2303.07205}, 2023.

\bibitem[Tourille et~al.(2022)Tourille, Sow, and
  Popescu]{tourille2022automatic}
Julien Tourille, Babacar Sow, and Adrian Popescu.
\newblock Automatic detection of bot-generated tweets.
\newblock In \emph{Proceedings of the 1st International Workshop on Multimedia
  AI against Disinformation}, pages 44--51, 2022.

\bibitem[Varol et~al.(2017)Varol, Ferrara, Davis, Menczer, and
  Flammini]{varol2017online}
Onur Varol, Emilio Ferrara, Clayton~A Davis, Filippo Menczer, and Alessandro
  Flammini.
\newblock Online human-bot interactions: Detection, estimation, and
  characterization.
\newblock In \emph{{Proceedings of the International AAAI Conference on Web and
  Social Media}}, 2017.

\bibitem[Wang et~al.(2023)Wang, Xie, Jiang, Mandlekar, Xiao, Zhu, Fan, and
  Anandkumar]{wang2023voyager}
Guanzhi Wang, Yuqi Xie, Yunfan Jiang, Ajay Mandlekar, Chaowei Xiao, Yuke Zhu,
  Linxi Fan, and Anima Anandkumar.
\newblock Voyager: An open-ended embodied agent with large language models.
\newblock \emph{Preprint arXiv:2305.16291}, 2023.

\bibitem[Weidinger et~al.(2022)Weidinger, Uesato, Rauh, Griffin, Huang, Mellor,
  Glaese, Cheng, Balle, Kasirzadeh, et~al.]{weidinger2022taxonomy}
Laura Weidinger, Jonathan Uesato, Maribeth Rauh, Conor Griffin, Po-Sen Huang,
  John Mellor, Amelia Glaese, Myra Cheng, Borja Balle, Atoosa Kasirzadeh,
  et~al.
\newblock Taxonomy of risks posed by language models.
\newblock In \emph{ACM Conference on Fairness, Accountability, and
  Transparency}, pages 214--229, 2022.

\bibitem[Yamin et~al.(2021)Yamin, Ullah, Ullah, and Katt]{yamin2021weaponized}
Muhammad~Mudassar Yamin, Mohib Ullah, Habib Ullah, and Basel Katt.
\newblock Weaponized ai for cyber attacks.
\newblock \emph{Journal of Information Security and Applications}, 57:\penalty0
  102722, 2021.

\bibitem[Yan et~al.(2022)Yan, Yang, Shanahan, and Menczer]{yan2022exposure}
Harry~Yaojun Yan, Kai-Cheng Yang, James Shanahan, and Filippo Menczer.
\newblock Exposure to social bots amplifies perceptual biases and regulation
  propensity.
\newblock \emph{SocArXiv}, 2022.

\bibitem[Yang et~al.(2023)Yang, Jin, Tang, Han, Feng, Jiang, Yin, and
  Hu]{yang2023harnessing}
Jingfeng Yang, Hongye Jin, Ruixiang Tang, Xiaotian Han, Qizhang Feng, Haoming
  Jiang, Bing Yin, and Xia Hu.
\newblock Harnessing the power of llms in practice: A survey on chatgpt and
  beyond.
\newblock \emph{Preprint arXiv:2304.13712}, 2023.

\bibitem[Yang et~al.(2019)Yang, Varol, Davis, Ferrara, Flammini, and
  Menczer]{yang2019arming}
Kai-Cheng Yang, Onur Varol, Clayton~A Davis, Emilio Ferrara, Alessandro
  Flammini, and Filippo Menczer.
\newblock Arming the public with artificial intelligence to counter social
  bots.
\newblock \emph{Human Behavior and Emerging Technologies}, 1\penalty0
  (1):\penalty0 48--61, 2019.

\bibitem[Yang et~al.(2020)Yang, Varol, Hui, and Menczer]{yang2020scalable}
Kai-Cheng Yang, Onur Varol, Pik-Mai Hui, and Filippo Menczer.
\newblock Scalable and {Generalizable} {Social} {Bot} {Detection} through
  {Data} {Selection}.
\newblock \emph{Proceedings of the AAAI Conference on Artificial Intelligence},
  34\penalty0 (01):\penalty0 1096--1103, 2020.
\newblock ISSN 2374-3468.

\bibitem[Yang et~al.(2022)Yang, Ferrara, and Menczer]{yang2022botometer}
Kai-Cheng Yang, Emilio Ferrara, and Filippo Menczer.
\newblock Botometer 101: Social bot practicum for computational social
  scientists.
\newblock \emph{Journal of Computational Social Science}, pages 1--18, 2022.

\bibitem[Ye et~al.(2023)Ye, Chen, Xu, Zu, Shao, Liu, Cui, Zhou, Gong, Shen,
  et~al.]{ye2023comprehensive}
Junjie Ye, Xuanting Chen, Nuo Xu, Can Zu, Zekai Shao, Shichun Liu, Yuhan Cui,
  Zeyang Zhou, Chao Gong, Yang Shen, et~al.
\newblock {A Comprehensive Capability Analysis of GPT-3 and GPT-3.5 Series
  Models}.
\newblock \emph{Preprint arXiv:2303.10420}, 2023.

\bibitem[Zhao et~al.(2023)Zhao, Wang, and Li]{zhao2023protecting}
Xuandong Zhao, Yu-Xiang Wang, and Lei Li.
\newblock Protecting language generation models via invisible watermarking.
\newblock \emph{Preprint arXiv:2302.03162}, 2023.

\bibitem[Zhou et~al.(2023)Zhou, Zhang, Luo, Parker, and
  De~Choudhury]{zhou2023synthetic}
Jiawei Zhou, Yixuan Zhang, Qianni Luo, Andrea~G Parker, and Munmun
  De~Choudhury.
\newblock Synthetic lies: Understanding ai-generated misinformation and
  evaluating algorithmic and human solutions.
\newblock In \emph{Proceedings of the CHI Conference on Human Factors in
  Computing Systems}, New York, NY, USA, 2023. ACM.

\end{thebibliography}

\end{document}